\documentclass[10pt, letterpaper]{IEEEtran}

\usepackage{subfig}
\usepackage{graphicx}
\usepackage{footnote}
\usepackage{color}
\usepackage{xcolor}
\usepackage{amsmath, amssymb}
\usepackage{stfloats}
\usepackage[normalem]{ulem}
\usepackage{epstopdf}
\usepackage[compress]{cite}
\usepackage[font={small}]{caption}
\usepackage{array}
\usepackage{bm}
\usepackage{booktabs}
\usepackage{braket}
\usepackage{url}
 \usepackage{multirow}
\usepackage{tikz}

\usepackage{mathtools}
\ifCLASSINFOpdf

\else

\fi

\IEEEoverridecommandlockouts
\usepackage[font={small}]{caption}

\begin{document}
	\bstctlcite{IEEEexample:BSTcontrol}
	\setlength{\parskip}{0pt}
	
	\title{Molecular Communication Theoretical Modeling and Analysis of SARS-CoV2 Transmission\\ in Human Respiratory System}
\author{Caglar Koca,~\IEEEmembership{Student Member,~IEEE}, Meltem Civas,~\IEEEmembership{Student Member,~IEEE}, \\Selin M. Sahin, Onder Ergonul, and Ozgur B. Akan,~\IEEEmembership{Fellow,~IEEE} \\
\thanks{C. Koca is with the Internet of Everything (IoE) Group, Electrical Engineering Division, Department of Engineering, University of Cambridge (e-mail: ck542@cam.ac.uk).}
\thanks{M. Civas is with Next-generation and Wireless Communications Laboratory (NWCL), Department of Electrical and Electronics Engineering, Ko\c{c} University, Istanbul, Turkey (e-mail: mcivas16@ku.edu.tr).}

\thanks{S. M. Sahin is with School of Medicine, Ko\c{c} University, Istanbul, Turkey (e-mail: ssahin20@ku.edu.tr).}

\thanks{O. Ergonul is with School of Medicine, Department of Infectious Diseases and Clinical Microbiology, Ko\c{c} University, Istanbul, Turkey and Ko\c{c} University Research Centre for Infectious Diseases, Istanbul, Turkey (e-mail: oergonul@ku.edu.tr).}

\thanks{O. B. Akan is with Next-generation and Wireless Communications Laboratory (NWCL), College of Engineering, Ko\c{c} University, Istanbul, Turkey; Internet of Everything Group, Department of Engineering, University of Cambridge, UK and Ko\c{c} University Research Centre for Infectious Diseases, Istanbul, Turkey and Koc University \  (e-mails: akan@ku.edu.tr, oba21@cam.ac.uk).} 

\thanks{This work was supported  in part by the AXA Research Fund (AXA Chair for Internet of Everything at Koc University), Huawei Graduate Research Scholarship and by Koc University \.{I}\c{s} Bank Research Center for Infectious Diseases (KUISCID)}}

\maketitle
\setcounter{page}{1} 

\begin{abstract}

		Severe Acute Respiratory Syndrome-CoronaVirus 2 (SARS-CoV2) caused the ongoing pandemic. This pandemic devastated the world by killing more than a million people, as of October 2020. It is imperative to understand the transmission dynamics of SARS-CoV2 so that novel and interdisciplinary prevention, diagnostic, and therapeutic techniques could be developed. In this work, we model and analyze the transmission of SARS-CoV2 through the human respiratory tract from a molecular communication perspective. We consider that virus diffusion occurs in the mucus layer so that the shape of the tract does not have a significant effect on the transmission. Hence, this model reduces the inherent complexity of the human respiratory system. We further provide the impulse response of SARS-CoV2-ACE2 receptor binding event to determine the proportion of the virus population reaching different regions of the respiratory tract. Our findings confirm the results in the experimental literature on higher mucus flow rate causing virus migration to the lower respiratory tract. These results are especially important to understand the effect of SARS-CoV2 on the different human populations at different ages who have different mucus flow rates and ACE2 receptor concentrations in the different regions of the respiratory tract.

\end{abstract}
\section{Introduction}
Information and communication technology (ICT) framework provides a novel perspective to and understand and fight human diseases \cite{akan2020information,civas2020rate,civas2018information, koca2017anarchy}. In this respect, molecular communication could pave the way for a solution to develop therapeutic and diagnostic platforms. Recent Severe Acute Respiratory Syndrome-CoronaVirus 2 (SARS-CoV2) pandemic have resulted in a significant number of deaths and adversely affected the whole humankind. Furthermore, an effective vaccine has not yet been developed. Molecular communication abstraction and characterization of the propagation of infectious diseases can provide new insight about these diseases exemplified by the viral infectious COVID-19 disease. 

Recent research  focused on the channel characterization of the virus infection and the transport of virus particles through aerosol transmission channel \cite{pal2019vivo,khalid2020modeling}. In \cite{pal2019vivo}, the authors model Dengue virus transmission inside the body from its entrance to the host to the transmission to affected organs. The channel considered, which is from skin to the receiver organs, is characterized in terms of noise sources and path loss. Aerosol transmission, in which droplets carry virus, is the another means of virus transport mechanism. In  \cite{khalid2020modeling}, the authors determine the aerosol channel impulse response and find the response of their system for the sources such as breathing, coughing and sneezing. On the other hand, a study considering the SARS-CoV2 transmission process through the human respiratory tract from molecular communication perspective is yet to be studied in the literature. 

\begin{figure}
	\centering

	\includegraphics[width=0.5\textwidth]{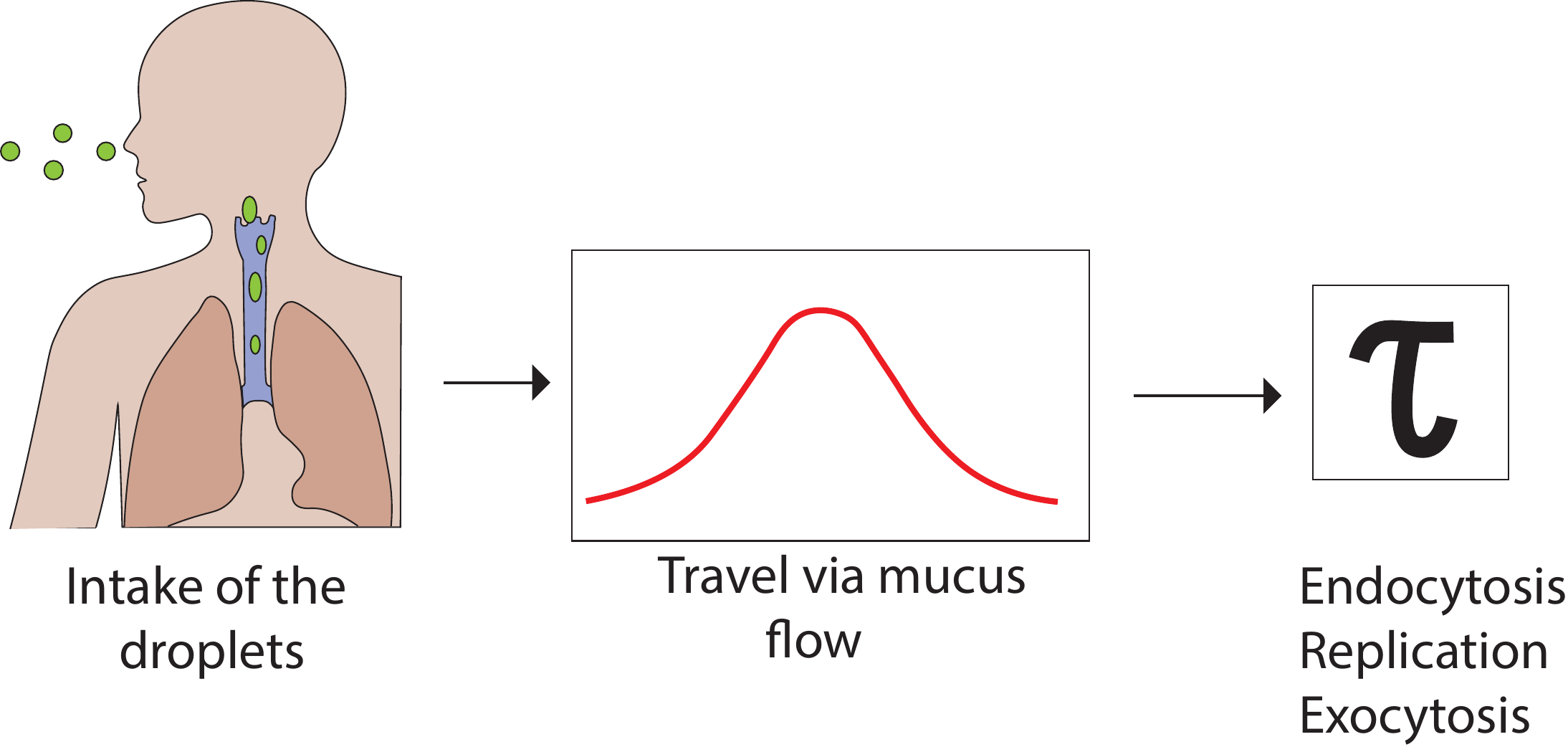}
	\caption{Overview of SARS-CoV2 transmission.}
	\label{fig:system}
\end{figure}

SARS-CoV2 enters the host human through the nose, mouth and eyes. We consider the case that droplets carrying viruses enter the host human from the nose. Viruses travel via mucus flow in the respiratory tract and reach host cells as illustrated in Fig. \ref{fig:system}. SARS-CoV2 virus binds a special receptor on the host cell called {\it angiotensin-converting enzyme} or ACE2, with a rate, $\lambda$. Binding is followed by a time delay, $\tau$, which is due to the mechanisms needed for virus replication. In this study, we consider this system and accordingly develop a model for the human respiratory tract by separating the respiratory tract into seven segments. Our aim is to determine the impulse response of the SARS-CoV2-ACE2 binding process to investigate the probability distribution of binding locations. The binding location distribution, which depends on several system parameters including ACE2 density and mucus flow rate offers vital information on the course of disease.

Our contributions can be summarized as follows:

\begin{itemize}
	\item {\bf \em Proposing a novel model of human respiratory tract that reduces complexities of the original system:} We model human respiratory tract by partitioning the tract into seven segments from nasal cavity to alveoli.    
	\item {\bf \em Calculating ACE2 receptor densities in the different regions of the respiratory tract:} Based on the available data on surface parameters, we calculate ACE2 receptor density crudely. 
	\item {\bf \em Determining impulse response of SARS-CoV2 infection process for the first time in literature:} We develop an impulse response for the propagation of SARS-CoV2 viruses in the respiratory tract, which is instrumental in further investigation of factor effecting disease progression.
	\item {\bf \em Investigating the effects of mucus layer thickness, mucus flow rate and ACE2 density on the virus population reaching the different regions of the respiratory tract and disease progression:} Our results show that mucus flow rate and ACE2 densities drastically affect the regions of respiratory tract where viruses reach.  
\end{itemize}

The rest of the paper is organized as follows. In Section \ref{sec:background}, we provide a brief background about SARS-CoV2. In Section \ref{sec:sysmodel}, the developed system model is outlined. In Section \ref{sec:diff}, the diffusion model for viruses diffusing through the mucus layer is derived. Next, in Section \ref{impulse_sec}, the impulse response of the system for different receptor and virus concentration is determined. In Section \ref{host}, Markov Chain model of the events following the binding process are stated. In Section \ref{perfo}, the simulation results are presented. Finally, conclusions are stated in Section \ref{concl}.

\section{Background}
\label{sec:background}
Severe Acute Respiratory Syndrome - CoronaVirus 2 (SARS-CoV2), also named novel-coronavirus (2019-n-Cov), has been identified as the causative infectious agent of coronavirus disease-19 (Covid-19), responsible for the current pandemic. Covid-19 has turned from a local pneumonia outbreak, which originated in Wuhan, China in December 2019, into a global pandemic in a matter of months, which has as of now, October 2020, caused more than a million deaths worldwide and spread to more than 200 countries \cite{coronavirus}. Belonging to the family of coronaviruses, SARS-CoV2 is the third and the newest coronavirus in the family to cause an epidemic, just as SARS-CoV in 2003 and MERS-CoV in 2012, and the only one to cause a pandemic. SARS-CoV2 is reported to be a zoonotic viral disease. Bats, snakes, and pangolins have been cited as potential reservoirs based on genome sequencing studies \cite{zhou2020pneumonia,zhang2020probable,ji2020cross}.

\subsection{Clinical Presentation}
Although it predominantly causes pneumonia and associated comorbidities, Covid-19 is considered to be a syndrome, given that it affects multiple different organs and systems within the human body. Typical clinical symptoms of the patients include fever, dry cough, difficulty of breathing (dyspnea), fatigue, joint pain (arthralgia), muscle pain (myalgia), and loss of sense of smell (anosmia) \cite{xu2020pathological,guan2020clinical,wolfel2020virological}. 
The presence of high variety of pathological events are attributed to different pathophysiological mechanisms involved in SARS-CoV2 and proves that it is more than a respiratory syndrome.

\subsection{Transmission Route}
Current epidemiological data suggests that SARS-CoV2 is an airborne viral disease, meaning that it is transmitted through respiratory droplets and droplet nuclei, which are mostly spread during human-to-human contact \cite{morawska2020airborne,liu2020community,li2020early}. Respiratory droplets ($>5-10 \mu$m in diameter) and droplet nuclei (aerosols) ($<5 \mu$m in diameter ), are generated and expelled/disseminated from an infected person during speaking, shouting, coughing, or sneezing \cite{world2014infection}. Indirect surface transmission, i.e., fomite transmission, and orofecal transmission have also been reported \cite{van2020aerosol,pastorino2020prolonged,heneghan2020sars}. Some studies have detected stable SARS-CoV2 viral RNA on solid surfaces such as plastic, aluminum, and stainless steel, yet the significance of fomite transmission is still debated with contradicting views \cite{van2020aerosol,pastorino2020prolonged}. 

The main pathway of SARS-CoV2 inside the human host is reported to be the respiratory tract. Mucosal openings such as the nose, eyes, or mouth have been identified as the principal sites, where the initial viral entry takes place \cite{gao2016sars}. Although there are numerous possibilities for viral entry, one pathway a virus particle could take on the macroscopic level is as follows. A virus laden particle enters through the nasal cavity, with the help of the downstream flow of mucosal secretions and gravity, it travels down through the pharynx, larynx, and trachea, enters a bronchi, passes down to bronchioles and finally reaches alveoli. On a microscopic level, once the virus laden droplets reach mucosal membranes, they diffuse through the mucosa (consisting of mucus, periciliary layer, and mid-layer) and attach to certain membrane receptors on host cell surfaces, the most prominent one being ACE2, which has been identified as the primary functional receptor for SARS-CoV2, just as for SARS-CoV \cite{li2003angiotensin,gu2007pathology,walls2020structure,zhao2020single}.

The current knowledge on SARS-CoV2 infection indicates that the elderly are more susceptible and vulnerable to the infection, while children seem to be the least affected group. Numerous studies report lower rates of SARS-COV2 infection with milder symptoms in children compared to adults \cite{covid2020coronavirus,castagnoli2020severe,rajapakse2020human}. Some studies attribute these results to their findings that ACE2 expression in children is lower than that of adults. \cite{bunyavanich2020nasal,wang2020single,sharif2020expression}. Other possible reasons held responsible for lower rates of morbidity and mortality from SARS-COV2 in children include: the differences in immune responses between children and adults, differences in ACE2 receptor distribution patterns, and lower rates of testing in children due to abundance of asymptomatic cases \cite{williams2020sars}.

\subsection{Viral Structure and Viral Binding}
The morphological structure of the virus comes to prominence when discussing viral binding processes. SARS-CoV2 is an enveloped, positive-sense, single-stranded RNA virus and similar to its prior relatives SARS-CoV and MERS-CoV, it belongs to Betacoronavirus genus of the coronavirinae family. 
SARS-CoV2 viral genome contains four major structural proteins: the \textit{S} (spike) protein, the \textit{M} (membrane) protein, the \textit{N} (nucleocapsid) protein, and the \textit{E} (envelope) protein \cite{schoeman2019coronavirus}. The \textit{S} protein has a trimeric structure, consisting of an \textit{S1} receptor binding subunit and an \textit{S2} fusion subunit. During viral infection, \textit{S1} and \textit{S2} subunits are cleaved by a metalloprotease, TMPRSS-2 (transmembrane protease serine 2), which facilitates viral entry. The \textit{S1} subunit functions as the part, which directly binds to the host cell receptor, i.e., ACE2 receptor, creating a Receptor Binding Domain (RBD). The \textit{S2} subunit takes role in membrane fusion \cite{yan2020structural}.

Following viral binding, there are two possible pathways of viral entry for enveloped viruses into host cells: either cytoplasmic fusion in which their envelope fuses with plasma membrane and they release their genome into cytosol, or endosomal membrane fusion (endocytosis) in which they are engulfed by an endosome, and their membrane is fused with the endosomal membrane \cite{white2016fusion,yang2020targeting}. 
There are multiple mechanisms of endocytic entry suggested by various studies, involving clathrin dependent, caveolae dependent endocytosis \cite{wang2008sars,milewska2018entry}, and clathrin independent, caveolae independent endocytosis \cite{inoue2007clathrin,glebov2020understanding}.

\section{System Model}
\label{sec:sysmodel}
In Section \ref{sec:background}, we present physio-morphological structure and behavior of the virus, regarding its entry mechanisms into human body and target cells. Here, we present our system model.

We assume that the virus carrying aerosol particles enter the human host
through the nose, and diffuse through the mucus layer in the nasal cavity, where ACE2 receptors
are found most abundantly \cite{sungnak2020sars}. The diffusion of the virus takes place in the mucus layer, which
renders the shape of the respiratory tract insignificant. Given the fact that the mucus layer is continuous within the respiratory tract \cite{hall2010guyton}, we
assume a cylindrical tube with radius $r(y)$ and length $l$. The change in the radius throughout the tract
has limited effect, unless it also modulates the properties of mucus, especially the mucus
thickness.

We know that the penetration of the virus containing droplets depend on the size of the droplet, with smaller droplets being able to penetrate further into the respiratory tract\cite{mucus}. Without loss of generality, we assume that the droplets are caught in the entrance of the   respiratory tract by the mucus layer due to the lack of information on the droplet size.

\begin{figure}[t]
	\centering
	\includegraphics[width=0.5\textwidth]{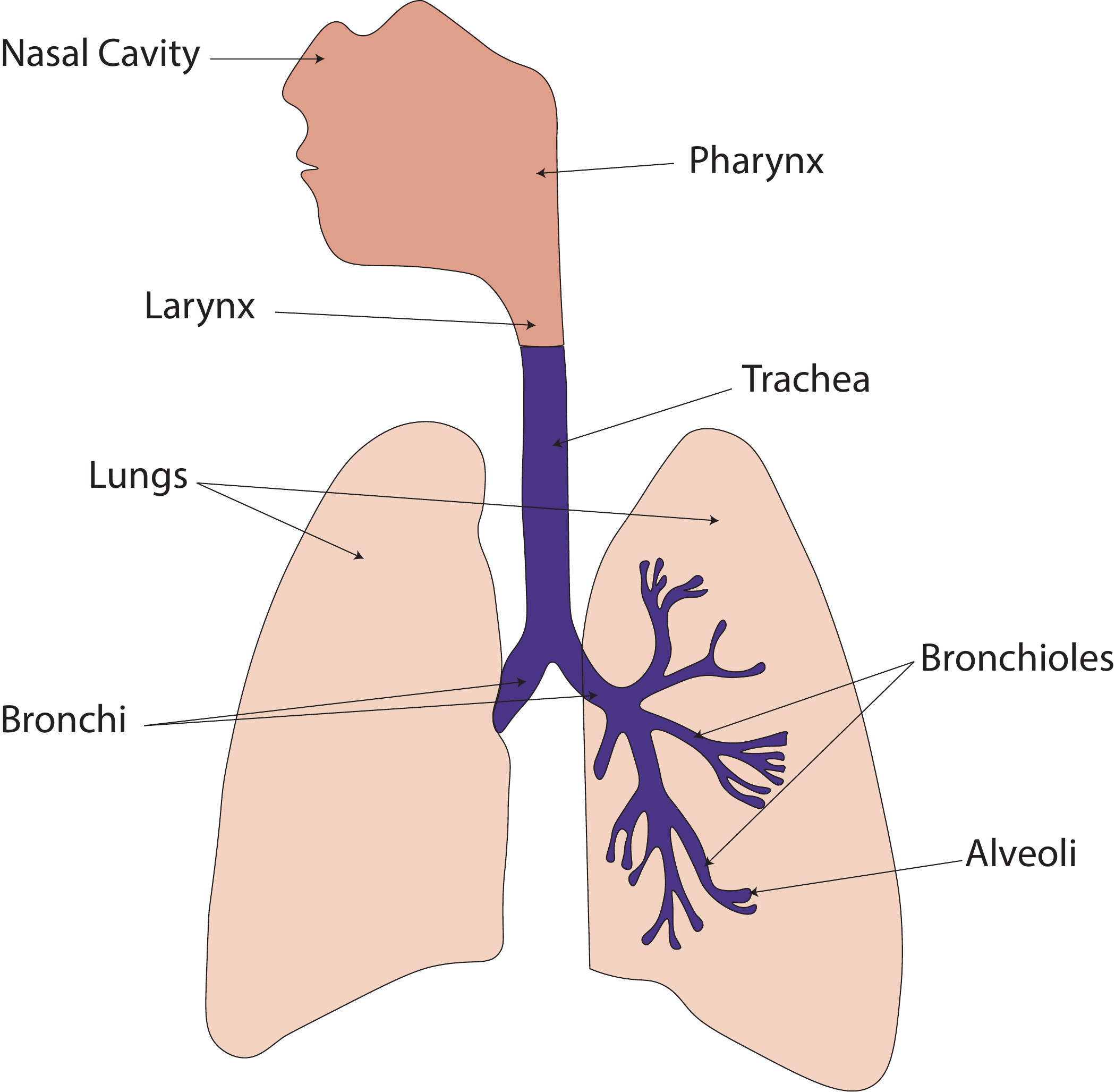}
	\caption{The human respiratory tract, separated into seven segments.}
	\label{fig:system2}
\end{figure}

\begin{figure}[t]
	\centering
	\includegraphics[scale=0.6]{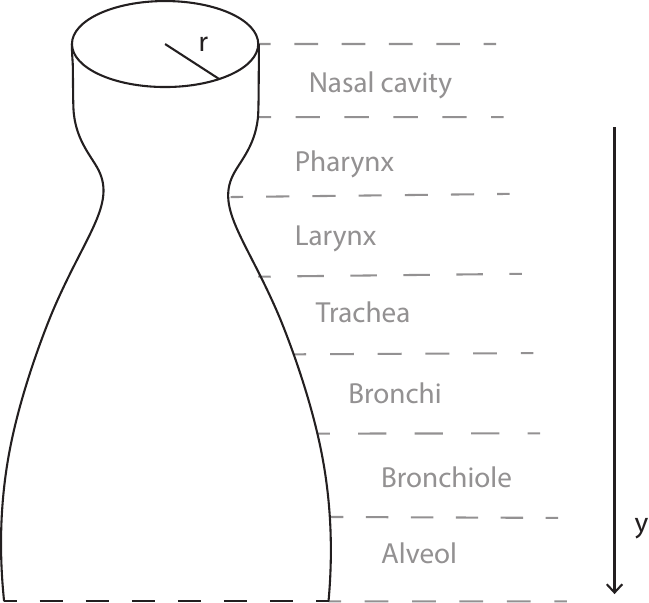}
	\caption{Simplified model for the human respiratory system.}
	\label{fig:system3}
\end{figure}

For a large portion of the respiratory tract, mucus layer covers the periciliary layer and a
thin surfactant layer separates the two \cite{rubin2002physiology}. In our work, we assume that the surfactant layer reduces the surface tension between these two layers to a negligible value, and consequently ignored. Furthermore, we assume that the diffusion coefficient, $D$, of the virus in periciliary and mucus layers to be the same. In a healthy respiratory system, the mucus inflow to the alveoli is countered by the mucus outflow due to the periciliary layer. We ignored the mucus outflow mechanism as it may turn the mucus flow into a very complex turbulent fluid model. In other words, we treat it as if it is one single layer.  

The existing works studying ACE2 distribution and mucus flow do not comment on differentiations within a region, i.e., ACE2 are homogeneously distributed. Hence, our model assumes cylindrical symmetry.

The virus moves under the influence of the mucus flow from nasal cavity to the alveoli\cite{mucus, mucus2}. We partition the respiratory system into seven parts, namely {\it Nasal Cavity, Larynx, Pharynx, Trachea, Bronchi, Bronchiole and Alveoli}. Our model is presented in Fig. \ref{fig:system2}. Because of the complicated structure of the bronchial tree, we assign transition regions to the closest region. Following bronchi, the respiratory tract branches out 22 more times. Due to this continuous branching in the lungs, we assign the sections where oxygen exchange occurs to alveoli and sections where no oxygen exchange occurs to bronchiole, meaning first 18 branches are assigned to bronchiole and last four to alveoli.\cite{gehrs_annexe} Furthermore, since after each branching, the individual branches become narrower but more numerous, we used the surface area, $S_i$, of each of the seven regions, $i \in \{1,2,...,7\}$, to calculate its corresponding radii values, $r_i$ as
\begin{equation}
\label{eq:radii}
r_i=\frac{S_i}{2\pi l_i},
\end{equation}
where $l_i$ is the length of the $i^{th}$ region. The resultant the respiratory tract, obtained by stacking cylinders, each with the radius and height of its corresponding segment, is shown in Fig. \ref{fig:system3}. Note that Fig. \ref{fig:system3} is not to scale, as the corresponding radii for alveol region is two orders larger than the next region, i.e., bronchiole region. Due to the cylindrical symmetry assumption, we can make a longitudinal cut through any point on the model and unwrap the respiratory tract.

Upon entering the mucus and periciliary layer, viruses use their viral S-spike proteins to bind to
ACE2 receptors on host cell surfaces \cite{guo2020origin}. We will use the binding rate, $\lambda$, to
describe the binding process. Due to the spherical shape of the coronavirus, we safely ignore the effect of the
orientation of the virus at the time when it makes contact with the ACE2 receptor.

As viruses bind to ACE2 receptors on the host cell’s membrane surface, ACE2 receptors are
downregulated. Therefore, the number of surface receptors decreases \cite{silhol2020downregulation,datta2020sars}, making it less likely for other viruses to bind. 

We consider two scenarios depending on the ACE2 receptor concentration and the virus population:
\begin{itemize}
	\item {\em Large virus and large receptor concentration:} The number of viruses and ACE2 receptors are the same at the locations $y$ and $y+\mathrm{d}y$.
	\item {\em Only large virus concentration:} The viruses dominate ACE2 receptors, i.e., all ACE2 receptors are bound to a virus and concurrently downregulated.
\end{itemize}
\begin{figure}[t]
	\centering
	\includegraphics[width=0.5\textwidth]{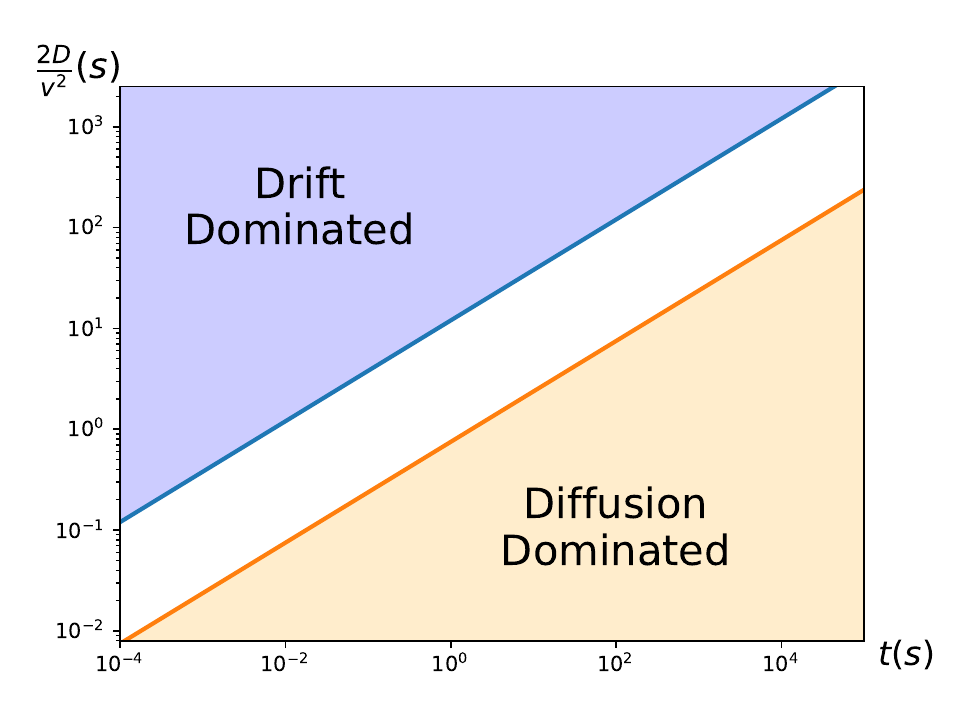}
	\caption{The dominant actors in virus movement, depending on mucus flow velocity, $v$, and diffusion coefficient, $D$, vs. time.}
	\label{fig:assumption}
\end{figure}

Following the ACE2 -- virus binding, endocytosis process occurs. We assume a constant endocytosis rate,  $r_{BE}$. As a result, our system model do not include the effects of caveolae and clathrin dependencies due to the lack of quantitative data. Once inside the cell, the virus releases its RNA and undergoes replication. The virus or its RNA may degrade before during and after replication, blocking the exocytosis and ending the infection of the cell.

\section{Viral Diffusion Model in the Respiratory Tract}
\label{sec:diff}
As stated in Section \ref{sec:sysmodel} we assume a constant mucus flow rate, $v$ from the nasal cavity to pharynx, larynx, trachea, bronchi, bronchiole and finally to the alveoli. Furthermore, the viruses also diffuse with a diffusion coefficient, $D$, in the mucus layer. The virus concentration is derived using Brownian Motion with drift. We assign $y$ axis for the distance from the entrance of nasal cavity, $x$ axis for the distance from a longitudinal cutting point and $z$ axis as the depth in the mucus layer. Due to the assumption of cylindrical symmetry, the reference point for $x$ coordinate is arbitrary. If a droplet containing $N$ viruses is incident to the mucus level at the location $(x_0, y_0, z_0)$, the virus concentration at time $t$ is

\begin{align}
C(x,y &,z, t)=\frac{N}{(4\pi{D}t)^(3/2)}\exp\left(\frac{-(x-x_0)^2}{4Dt}\right)\\ &\exp\left(\frac{-(y-y_0+vt)^2}{4Dt}\right)\exp\left(\frac{-(z-z_0)^2}{4Dt}\right). \nonumber
\end{align}

\begin{figure}[t]
	\centering
	\hspace*{-3.5mm}	   			\vspace*{-2.5cm}	\includegraphics[width=0.55\textwidth]{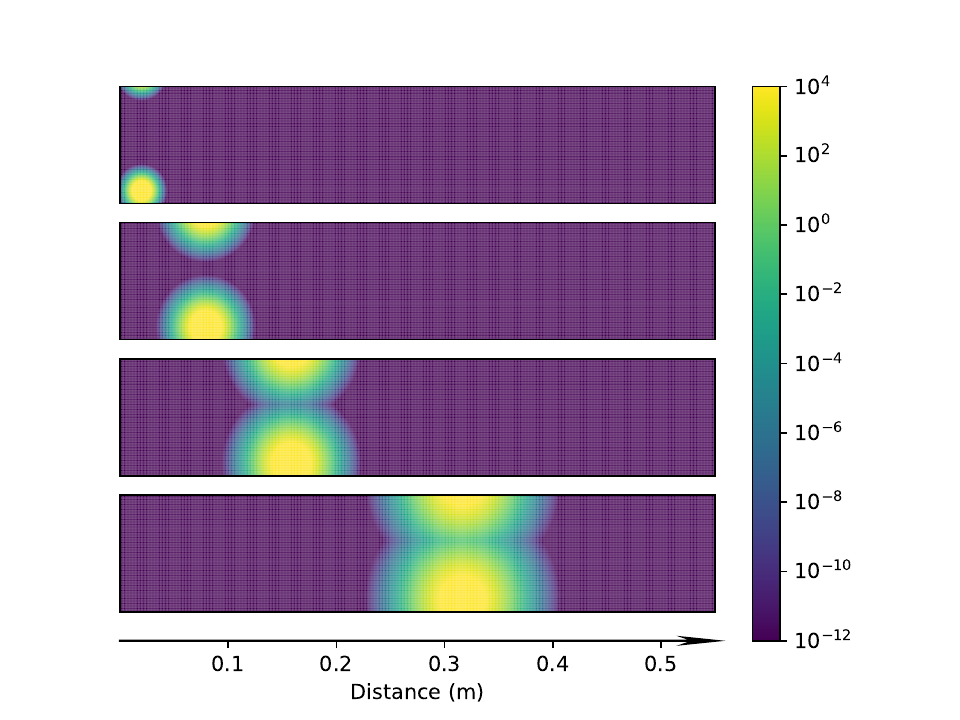}
	\vspace*{2.5cm}
	\caption{The diffusion of the virus in the respiratory tract for $t=0.5h$, $t=2h$, $t=4h$, $t=8h$ from top to bottom. The unit of the density is $m^{-3}$.}
	\label{fig:4fig}
\end{figure}

The standard deviation for Brownian motion is given as $\sigma=\sqrt{2Dt}$. Therefore, $95.45\%$ of the population of viruses falls into a sphere of radius $2\sqrt{2Dt}$, centred at $(x_0, y_0+vt,z_0)$, while $99.7\%$ into a sphere with the same centre and radius of $3\sqrt{2Dt}$. Hence, for $vt \gg 3\sqrt{2Dt}$, drift dominates the diffusion and diffusion along the y-axis can be ignored. Similarly, for $vt \ll 3\sqrt{2Dt}$, drift is dwarfed by diffusion and can be ignored. Fig. \ref{fig:assumption} shows the dominating trends for Brownian Motion with drift. For $v=50\mu {m}s^{-1}$ \cite{chen1978mucus} and $D=1.5\times 10^{-11} m^2s^{-1}$ \cite{block2016quantification,Fain2020,Diff20200}, the diffusion of the virus in the respiratory tract is shown in Fig. \ref{fig:4fig}. Clearly, the effects of the diffusion is only visible for large $t$.

\bgroup
\def\arraystretch{1.5}
\begin{table}[]
	\centering
	\begin{tabular}{|l|c|}
		\hline
		\textbf{Variable}                        & \textbf{Symbol} \\ \hline
		{\it Concentration}                            & $C$             \\ \hline
		{\it Concentration due to a single virus}      & $C_1$           \\ \hline
		{\it ACE2--Virus Binding Rate}                 & $\lambda$       \\ \hline
		{\it Distance from the start of the tract}     & $y$             \\ \hline
		{\it Initial number of viruses}					 & $N_0$		   \\ \hline
		{\it Virus concentration} 					 & $N(y)$		   \\ \hline
		{\it Tract radius}                             & $r(y)$          \\ \hline
		{\it Height of mucus}                          & $h(y)$          \\ \hline
		{\it Mucus flow rate}                          & $v$             \\ \hline
		{\it Avogadro's Constant}                      & $N_A$           \\ \hline
		{\it Binding probability to a single ACE2}     & $p_b$           \\ \hline
		{\it Not binding probability to a single ACE2} & $p_{nb}$        \\ \hline
		{\it Evade probability in the segment}         & $p_e(y)$           \\ \hline
		{\it ACE2 concentration}                       & $f(y)$          \\ \hline
		{\it ACE2 count}                               & $n(y)$          \\ \hline
		{\it Number of viruses at $y$}                              & $N(y)$          \\ \hline
		{\it Probability of reaching $y$}              & $p(y)$          \\ \hline
		{\it Binding rate at $y$}            & $p_b(y)$        \\ \hline
		{\it Virus distribution}                       & $V(y)$          \\ \hline
		{\it Unbinded (free) viruses}                  & $\bar{V}(y)$    \\ \hline
		{\it Expected number of bindings}              & $E_b$           \\ \hline
	\end{tabular}
	\caption{List of symbols used in the analysis}
\end{table}
\egroup
%
%
%

\section{Impulse Response of Virus-ACE2 Binding}
\label{impulse_sec}
The ACE2-Virus binding can be modelled by obtaining the virus population distribution over the respiratory tract. To achieve this, we start with modelling the kinematics of a single virus incident on the mucus layer. Later, we use our findings as stepping stones to reach impulse response for different scenarios as described in Section \ref{sec:sysmodel}.

\subsection{Kinematics of Single Virus Case}

We begin our analysis by considering a single virus is moving under the influence of mucus flow. The mucus layer has a thickness of $h(y)$ and a velocity of $v$, while the respiratory tract radius is $r(y)$, where $y$ lies in the direction of the respiratory tract from nose to lungs. Then, at any segment $\mathrm{d}y$, the concentration of the virus due to a single virus is given by
\begin{equation}
\label{eq:conc}
C_1=\frac{1}{2\pi{r(y)}h(y)\Delta y}.
\end{equation}

The time $\Delta t$ that the virus spends in a segment of length $\Delta y$ is 
\begin{equation}
{\Delta}t=\frac{{\Delta}y}{v}.
\end{equation}

The probability that it binds to a single ACE2 receptor in the segment with length ${\Delta}y$ becomes
\begin{align}
p_b&=1-\exp(-\lambda^1{C_1}{\Delta}t) \\
&=1-\exp\left(-\lambda^1\frac{1}{2\pi{r(y)h(y){\Delta}y}}\frac{{\Delta}y}{v}\right)\\
&=1-\exp\left(-\frac{\lambda}{N_A\pi{r(y)h(y)v}}\right) \\
\label{eq:pbb}
&=\frac{\lambda}{N_A\pi{r(y)h(y)v}},
\end{align}
where $\lambda$ is the molar association constant, $N_A$ is the Avogadro's constant and $\lambda^1=\lambda/N_A$ is the association constant for a single virus. Note that in the last step, we used first order Taylor series expansion, i.e., $e^x=1+x$ for small $x$.

Then, $p_{nb}$, the probability of not binding during ${\Delta}t$ is
\begin{align}
p_{nb}&=1-p_b \\
&=1-\frac{\lambda}{N_A\pi{r(y)h(y)v}}.
\end{align}

If the ACE2 concentration per unit area at $y$ is $f(y)$, then number of ACE2 receptors, $n(y)$, in the patch of length ${\Delta}y$ becomes
\begin{equation}
\label{eq:ny}
n(y)=2\pi{r(y)f(y){\Delta}y},
\end{equation}
and the probability of the virus evading all ACE2 receptors in the same patch, $p_e(y)$, is expressed as
\begin{align}
p_e(y)&=p_{nb}^{n(y)} \\
\label{eq:binom1}
&=\left(1-\frac{\lambda}{N_A\pi{r(y)h(y)v}}\right)^{2\pi{r(y)f(y){\Delta}y}} \\
\label{eq:binom2}
&=1-\frac{\lambda 2\pi{r(y)f(y){\Delta}y}}{N_A\pi{r(y)h(y)v}} \\
\label{eq:binom3}
&=1-\frac{2\lambda f(y){\Delta}y}{N_Ah(y)v},
\end{align}
where from \eqref{eq:binom1} to \eqref{eq:binom2} we use the first order truncation of the binomial expansion, i.e., $(1+x)^n=1+nx$ for $\lvert nx \rvert \ll 1$, which holds due to $N_A$ being much larger than any other value in \eqref{eq:binom2}. This assumption is especially effective for $\Delta{y}\rightarrow \mathrm{d}y$.

From \eqref{eq:binom3}, we reach the rate of binding in the patch of length $\mathrm{d}y$ as
\begin{align}
p_b(y)&=1-p_e(y) \\
\label{eq:pbbb}
&=\frac{2\lambda f(y)\mathrm{d}y}{N_Ah(y)v}.
\end{align}

Then, we find to the number of viruses at $y$, $N(y)$ using an initial value problem with rate $p_b(y)$

\begin{align}
\mathrm{d}N(y) &=-N(y)\frac{2\lambda f(y)\mathrm{d}y}{N_A h(y)v}\\
\mathrm{d}N(vt)&=-N(vt)\frac{2\lambda f(vt)\mathrm{d}t}{N_A h(vt)}\\
\frac{\mathrm{d}N(vt)}{N(vt)} &=-\frac{2\lambda f(vt)}{N_A h(vt)}\mathrm{d}t \\
\label{eq:nyyy}
N(vt)&=N_0\exp\left[-\int_0^t \frac{2\lambda f(vt^\prime)\mathrm{d}t^\prime}{N_A h(vt^\prime)} \right],
\end{align} 
where we used the fact that $\mathrm{d}y=v\mathrm{d}t$ by definition.

An important observation is that $N(y)$ obtained in \eqref{eq:nyyy} does not necessarily normalise. Defining $\bar{V}$ as 
\begin{equation}
\bar{V}\triangleq 1-\int_0^lN(y)\mathrm{d}y,
\end{equation}
where $l$ is the total length of the respiratory tract, $\bar{V}$ gives us the rate of the viruses that reach the end of the respiratory tract, i.e., alveoli. Here, since the viruses cannot travel neither forward nor backward, we assume that they will eventually bind to an ACE2 in alveoli.

\subsection{Modelling of ACE2-Virus Binding}

\begin{figure}[t]
	\centering
	\includegraphics[width=0.5\textwidth]{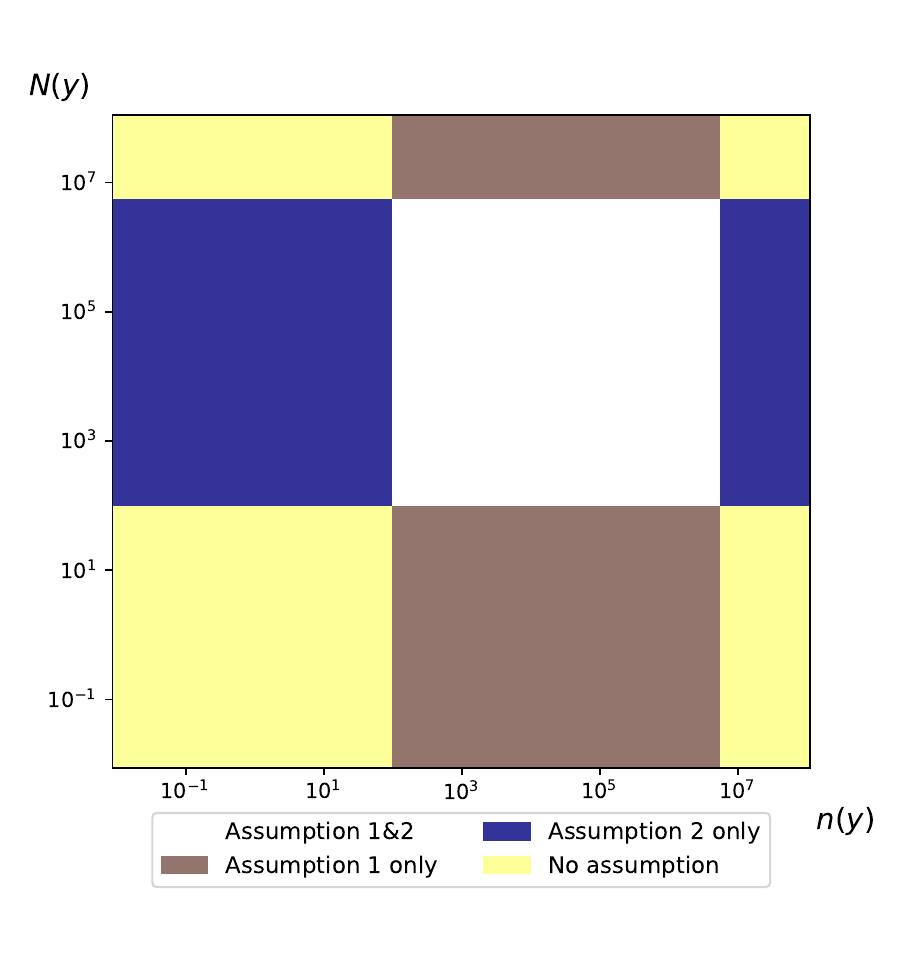}
	\caption{The regions where the constant $N(y)$ and constant $n(y)$ assumptions hold for $A(y)\sim 10^{-10}$.}
	\label{fig:Nn_ass}
\end{figure}
So far, we only assumed the existence of a single virus to reach \eqref{eq:nyyy}. As stated in Section \ref{sec:sysmodel}, there are several scenarios depending on 
\begin{itemize}
	\item $N(y)=2C(y)\pi r(y)h(y)\mathrm{d}y $, the total number of viruses number on a $\mathrm{d}y$ thick strip,
	\item $n(y)=2\pi r(y)f(y)\mathrm{d}y$, total number of ACE2 receptors in the same strip,
	\item $E_b$, the expected number of virus bindings in the same strip.
\end{itemize}

$E_b$ is loosely calculated by replacing $C_1$ with $C$ and carrying out \eqref{eq:conc} to \eqref{eq:binom3}. Hence, \eqref{eq:conc} becomes
\begin{equation}
C=\frac{N(y)}{2\pi r(y) h(y)\mathrm{d}y},
\end{equation}
and replacing $f(y)$ with $n(y)/2\pi r(y)\mathrm{d}y$,
\begin{align}
E_b &\simeq \frac{\lambda}{N_A\pi h(y)r(y)v}N(y)n(y) \\
\label{eq:eb}
& \simeq A(y) N(y)n(y),
\end{align}
where $A(y)$ does not depend on ACE2--virus bindings.

Since each ACE2--virus binding destroys both a virus and a receptor, both the virus and the receptor concentrations are affected. Hence, our model must incorporate variations in the concentrations. Each binding causes the number of viruses, $N(y)$ and number of receptors$n(y)$ to change as

\begin{align}
\label{eq:NYup}
N(y) &\longleftarrow N(y)-1 \\
\label{eq:ny_up}
n(y) &\longleftarrow n(y)-1
\end{align}

Using \eqref{eq:eb}, \eqref{eq:NYup} and \eqref{eq:ny_up} we reach two assumptions:

\begin{enumerate}
	\item \textit{Large $N(y)$:} If $N(y)$ is large and $E_b \ll N(y)$, the total virus concentration remains constant within the same segment.
	
	\item \textit{Large $n(y)$:} If $n(y)$ is large and $E_b \ll n(y)$, the total ACE2 concentration remains constant within the same segment.
\end{enumerate}

Since $A(y)$ is quite low, i.e., on the order of $10^{-10}$, for some cases, both of these assumptions hold. Fig. \ref{fig:Nn_ass} illustrates under which conditions these assumptions hold.

As Fig. \ref{fig:Nn_ass} shows, when both $N(y)$ and $n(y)$ are large, the assumptions may not hold. Since $A(y)$ depends on $r(y)$ and $h(y)$, the boundaries may change.

Note that changes in $n(y)$ causes a change in the system. As the system parameters change with the input, the system is no longer linear time-invariant. As a result, obtaining the impulse response when \eqref{eq:ny_up} does not hold is of no practical use.



\begin{figure*}

	\begin{center}
\begin{tikzpicture}
\node[draw] (A) at (0,0) {$B$};
\node[draw] (B) at (2.5,0) {$END$};
\node[draw] (C) at (5,0) {$C$};
\node[draw] (D) at (7.5,0) {$R$};
\node[draw] (E) at (10,0) {$EXC$};
\node[draw] (F) at (5,-3) {$D$};
\draw[thick,->] (A) -- (B)node[midway,above] {$r_{BE}$};
\draw[thick,->] (B) -- (C)node[midway,above] {$r_{EC}$};
\draw[thick,->] (C) -- (D)node[midway,above] {$r_{CR}$};
\draw[thick,->] (D) -- (E)node[midway,above] {$r_{RE}$};
\draw[thick,->] (B) -- (F)node[midway,left] {$r_{ED}$};
\draw[thick,->] (C) -- (F)node[midway,right] {$r_{CD}$};
\draw[thick,->] (D) -- (F)node[midway,right] {$r_{RD}$};
\end{tikzpicture}
	\caption{\color{black}{State transitions for life cycle of the virus in the host cell.}}
		\label{life_cycle}
\end{center}
\end{figure*}
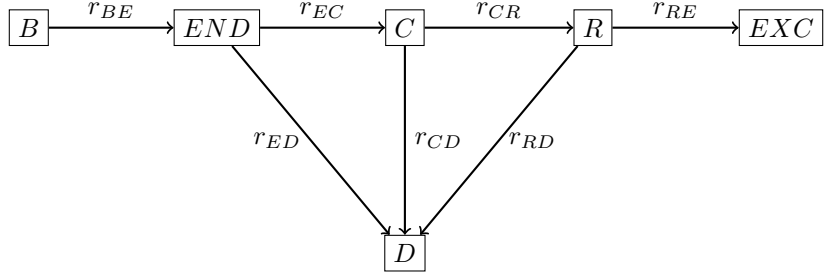

\subsubsection{Large $n(y)$}
For large $n(y)$, the virus--ACE2 bindings do not change the receptor number in the same segment. Hence, regardless of how many bindings happen in a given segment, the binding probability of any virus in the same segment is constant. As a result, \eqref{eq:pbbb}, multiplied by $N(y)$ gives the expected number of binding on the segment. Furthermore, \eqref{eq:nyyy} multiplied by the incident virus count, $N_0$, gives us the virus population reaching to any location $y$ of the respiratory tract, i.e., 

\begin{equation}
\label{eq:additional}
N(vt)=N_0\exp\left[-\int_0^t \frac{2\lambda f(vt^\prime)\mathrm{d}t^\prime}{N_A h(vt^\prime)} \right].
\end{equation}

Hence, the virus concentration is simply the derivative of \eqref{eq:additional}, i.e.,

\begin{equation}
V(y)=-\frac{\partial N(y)}{\partial y}
\end{equation}.

We proceed to obtain impulse response, $I(y, t)$, by adding the unbounded or free virus distribution, $\bar{V}(y)$. The free virus population is situated at the location $y=vt$ of the respiratory tract, due to the fact that virus movement on the respiratory tract is solely under the influence of mucus flow. The total number of free viruses is equal to the difference between the initial number of viruses and the total number of bound viruses.

\begin{align}
\bar{V}(y,t)=(N_0-V(vt))\delta(y-vt),
\end{align}
where $\delta(.)$ is the Dirac Delta function. Thus, the impulse response becomes
\begin{equation}
\label{eq:1Nn_imp}
I(y,t)=V(y,t)+\bar{V}(y,t).
\end{equation}

\subsubsection{Large $N(y)$ only}
In case $E_b \ll N(y)$, viruses outnumber the ACE2 receptors. This causes all ACE2 receptors to bind to a virus. Bound virus distribution in the respiratory tract is the same as the ACE2 receptor concentration. Note that, since bound ACE2 receptors downregulate, large $N(y)$ only case cannot be modelled as a linear time invariant system.

\begin{equation}
\label{eq:ACE2d}
n(y)=2\pi r(y)f(y)\mathrm{d}y,
\end{equation}
and
\begin{equation}
\label{eq:4nn1}
V(y,t)=n(y)\left[u(vt-y)-u(y)\right],
\end{equation}
where $u(.)$ is the step function and is used to assured that virus distribution is limited to the region $0-vt$, i.e., the range of mucus flow. Free virus distribution can be calculated by finding the number of receptors in the given region.

\begin{equation}
\label{eq:4nn2}
\bar{V}(y,t)=\left(N_0-\int_0^{vt}2\pi r(y) f(y)\mathrm{d}y\right)\delta(y-vt).
\end{equation}

We find the impulse response by adding \eqref{eq:4nn1} and \eqref{eq:4nn2}, i.e.,

\begin{align}
I(y,t)&=n(y)\left[u(vt-y)-u(y)\right]+\\ &\left(N_0-\int_0^{vt}2\pi r(y) f(y)\mathrm{d}y\right)\delta(y-vt)
\end{align}

\subsubsection{No Limiting Case}
In case both assumptions fail, active number of ACE2 receptors constantly changes due to the binding viruses. Therefore, no assumption can be made for this case, and neither \eqref{eq:1Nn_imp} nor \eqref{eq:4nn2} holds. As a result, there is no closed form expression for this case.

Note that, as described in Section \ref{sec:sysmodel}, we assume that the virus containing droplets are captured by the mucus layer at the start of the respiratory tract. The penetration of the virus laden droplets into the respiratory tract can be easily taken into account by finding the convolution of the droplet penetration distribution with the impulse response obtained here.

\section{Entry and Life Cycle in the Host Cell}
\label{host}
In the host cell, the virus replicate and new virions are released out of the cell via exocytosis. We can model this process with a stationary Markov Chain with six states, namely, binding ($B$), endocytosis ($END$), release of viral RNA ($C$), replication ($R$), degradation ($D$), exocytosis ($EXC$) as illustrated in Fig. \ref{life_cycle}.  
The bound virus, can enter the cell via endocytosis, which is mediated by ACE2 receptors. In the state-B, the virus is found bounded to the ACE2 receptor. COVID-19 is an RNA virus, i.e., virus can replicate in the cytoplasm. Thus, in the state-$C$ viral RNA is released to the cytoplasm. In the host cell, the virus can be degraded by lysosomes \cite{heldt2015single}, which is represented by the state-$D$.   

The transition matrix of the Markov Chain representing the life cycle of the virus in the host cell, $Q$, is given by 
\begin{equation}
\label{eqn_dbl_x}
Q = 
\begin{pmatrix}
G_1 & r_{BE} & 0 & 0 & 0 & 0\\
0 & G_2 & r_{EC} & 0 & r_{ED} & 0\\
0 & 0 & G_3 & r_{CR} & r_{CD} & 0 \\
0 & 0 & 0 & G_4  & r_{RD} & r_{RE} \\
0 & 0 & 0 & 0 & 1 & 0 \\
0 & 0 & 0 & 0 & 0 & 1\\
\end{pmatrix},
\end{equation}
where we set $G_{1} = -r_{BE}$, $ G_2 =  -(r_{EC} + r_{ED}) $, $G_3 = -(r_{CR} + r_{CD})$, and $G_4 = -(r_{RD} + r_{RE})$. The transition rates are provided in Table \ref{Table:Rates}. 


\begin{table}
	\def\arraystretch{1.3}
	\footnotesize
	\centering
	\caption{Rate parameters for Markov Chain representing virus life cycle in the host cell.}
	\begin{tabular}{|l|} \hline
		\textbf{Parameter} \\		\hline
		{\emph{$r_{BE}$ - Endocytosis rate}}  \\ 
		\hline
		{\emph{$r_{EC}$ - Viral RNA release rate}} \\ 
		\hline
		{\emph{$r_{ED} $ - Degradation rate}} \\ 
		\hline
		{\emph{$r_{CR} $ - Replication rate}} \\ 
		\hline
		{\emph{$r_{CD} $ - Degradation rate}} \\ 
		\hline
		{\emph{$r_{RD} $ - Degradation rate}} \\ 
		\hline
		{\emph{$r_{RE} $ - Exocytosis rate}} \\ 
		\hline
	\end{tabular}
	\label{Table:Rates}
	\vspace{-2mm}
\end{table}
$P(t) = [p_B(t), p_{END}(t), p_{C}(t), p_R(t), p_D(t), p_{EXC}(t)]$ represents corresponding state occupancy probabilities. The relation between the states can be expressed as\cite{eckford2015information}
\begin{equation}
\label{eqn_42}
\frac{\mathrm{d}P(t)}{\mathrm{d}t} = P(t) Q,
\end{equation}
which has a solution of \eqref{eqn_42} is in the form of 
\begin{equation}
P(t) = e^{Qt}.
\label{eqn_44}
\end{equation}
Using eigenvalue decomposition, we can express $Q$ as
\begin{equation}
Q = \sum_{i}^{}\lambda_i \vec{v}_i \vec{v}_i^T
\end{equation}
where $\vec{v}_i$ is an eigenvector of the matrix and $\lambda_i$ the corresponding eigenvalue. As a result, we can express \eqref{eqn_44} as 
\begin{equation}
e^{Qt} = \sum_{i}e^{\lambda_i t} \vec{v}_i \vec{v}_i^T.
\end{equation}
Thus, the probability of transition from the state $j$ to the state $k$ in time $t$ is given by 
\begin{equation}
P_{jk}(t) = P(S(t) = k|S(0) = j) = \sum_{i}e^{\lambda_i t} [\vec{v}_i \vec{v}_i^T]_{jk}.
\end{equation}

\section{Performance Evaluation}
\label{perfo}
\subsection{Physiological Parameters}
Morphometric measurements of the respiratory tract such as length, diameter, surface area, and mucosal thickness were obtained from estimation studies, journals, databases, and anatomy literature. Given the non-uniform shape and the continuously narrowing nature of the respiratory tract, as in the tracheobronchial tree, we use median values for branching or narrowing structures. Divided into 23 generations of dichotomous sections, the tracheobronchial tree designates a generation for each divided branch starting from \emph{trachea}, generation 0, and ending at \emph{alveoli}, (generation 23). The first 16 generations, from generation 0 to generation 16, are defined as \emph{the conducting zone}, i.e., no gas exchange takes place in this region. From generation 17 to generation 23 is called \emph{the transitional and respiratory zone}, where gas is exchanged within functional units \cite{patwa2015anatomy}. The generation 0 directly gives the dimensions for the trachea. Generations 1 to 4 are assumed to be \emph{bronchi}, 5 to 16, \emph{bronchiole} and 17 to 23 \emph{alveoli} respectively. \footnote{For our parameter data, we mainly use Gehr's ``Annexe A. Anatomy and Morphology of the Respiratory Tract'' and Weibel's ``Morphometry of the human lung''. \cite{gehrs_annexe,weibel1963}.}

Although there are studies investigating ACE2 receptor gene and protein expressions across different tissues and in specific cell types using single cell RNA sequencing gene expression profiling datasets, mass spectrometry and immune assay techniques, \cite{guo2020single,li2020expression,descamps2020ace2,zhao2020single}, to the best of our knowledge, data on the number of ACE2 receptors on different tissues is not explicitly stated in studies. Most studies provide relative expressions of the receptor in different tissues, shown as proportions, percentages, or plots with no numeric values. Some studies address circulating ACE2 levels, which we cannot directly utilize as we need tissue-specific values. There exist some studies which report ACE2 expression data in animals, which are not compatible with our work either \cite{xudong2006age}.

The primary challenge of this work is to obtain the ACE2 receptor densities in different tissues of respiratory tract. The lack of studies giving these values is mostly due to the difficulty of measuring ACE2 receptor concentration in a diverse population of all ages. To address this challenge, we exhaustively search among various literature to calculate our estimated values. The specific works that we use are referenced in Table III.

Therefore, we first gather data on the percentage of ACE2 expressing cells for the seven region model described in Section \ref{sec:sysmodel}. Then, we search for the total number of cells in each region. For tissues in which there is no sufficient quantitative data on the percentage of ACE2 expressing cells, the relative proportions of ACE2 expressions of two or more tissues, one of which we have previously calculated are used. Then, we calculate the number of ACE2 expressing cells in each tissue accordingly. 
Note that these preliminary calculations are the estimates based on the currently available data in the literature. 


Due to the lack of data, the effect of age in SARS-CoV2 susceptibility cannot be directly analyzed. However, we investigate the effects of thicker mucus as seen more in elderly and effects of higher ACE2 concentration in nasal cavity as observed in smokers.

\bgroup
\def\arraystretch{1.1}
\begin{table*}[t]
\centering
\begin{tabular}{c|c|c|c|c|l}
\textbf{}                                           & \textbf{Symbol}          & \textbf{Region}  & \textbf{Value}       & \textbf{Units}                                                & \textbf{Reference}    \\ 
\hline
\multirow{7}{*}{\textit{Region Length} }            & \multirow{7}{*}{$y$ }    & Nasal Cavity     & 11.2                 & \multirow{7}{*}{$cm$}                                       &\cite{gehrs_annexe}                       \\
                                                    &                          & Pharynx          & 14                   &                                                               &\cite{gehrs_annexe}                       \\
                                                    &                          & Larynx           & 4.4                  &                                                               &\cite{gehrs_annexe}                       \\
                                                    &                          & Trachea          & 12                   &                                                               &\cite{gehrs_annexe}                       \\
                                                    &                          & Bronchi          & 8.7                  &                                                               &\cite{gehrs_annexe}                       \\
                                                    &                          & Bronchiole       & 6.2                  &                                                               &\cite{gehrs_annexe}                       \\
                                                    &                          & Alveoli          & 0.36                 &                                                               &\cite{gehrs_annexe}                       \\ 
\hline
\multirow{7}{*}{\textit{Mucus Thickness} }          & \multirow{7}{*}{$h(y)$} & Nasal Cavity     & 2500                 & \multirow{7}{*}{$\mu{m}$}                  &                       \\
                                                    &                          & Pharynx          & 1600                 &                                                               &                       \\
                                                    &                          & Larynx           & 1600                 &                                                               &                       \\
                                                    &                          & Trachea          & 200                  &                                                              &\cite{gehrs_annexe}                       \\
                                                    &                          & Bronchi          & 60                   &                                                               &                      \cite{gehrs_annexe} \\
                                                    &                          & Bronchiole       & 6                    &                                                               &                      \cite{gehrs_annexe} \\
                                                    &                          & Alveoli          & 0.1                  &                                                               &                       \\ 
\hline
\multirow{7}{*}{\textit{ACE2 density} }             & \multirow{7}{*}{$f(y)$ } & Nasal Cavity     & 1000                 & \multirow{7}{*}{$10^9 \times m^{-2}$} &\cite{hou2020sars}                       \\
                                                    &                          & Pharynx          & 65                   &                                                               &                      \cite{hou2020sars} \\
                                                    &                          & Larynx           & 270                  &                                                               &                      \cite{hou2020sars} \\
                                                    &                          & Trachea          & 12                   &                                                               &                       \\
                                                    &                          & Bronchi          & 68                   &                                                               &                      \cite{mercer1994cell} \\
                                                    &                          & Bronchiole       & 18                   &                                                               &                      \cite{mercer1994cell} \\
                                                    &                          & Alveoli          & $1.5\times 10^{-7}$  &                                                               &                      \cite{crapo1982lung,braga2019cellular,reyfman2019single} \\ 
\hline
\multirow{7}{*}{\textit{Surface Area} }             & \multirow{7}{*}{$S(y)$ } & Nasal Cavity     & 160                  & \multirow{7}{*}{$cm^{2}$}                                  &\cite{kacmarek2013essentials, gizurarson2012anatomical} \\
                                                    &                          & Pharynx          & 50                   &                                                               &                      \cite{ali1965histology} \\
                                                    &                          & Larynx           & 50                   &                                                               &                      \cite{ali1965histology} \\
                                                    &                          & Trachea          & 70.83                &                                                               &                       \cite{hasan2007estimating} \\
                                                    &                          & Bronchi          & 40.10                &                                                               &                       \cite{hasan2007estimating} \\
                                                    &                          & Bronchiole       & 229                  &                                                               &                       \cite{hasan2007estimating} \\
                                                    &                          & Alveoli          & $1.43\times 10^6$    &                                                               &                       \cite{kacmarek2013essentials}\\ 
\hline
\multicolumn{1}{l|}{\textit{Mucus Flow Rate}}       & $v$                    & \multicolumn{2}{c|}{50}                 & $\mu{m}s^{-1}$                        &\cite{chen1978mucus}                       \\ 
\hline
\multicolumn{1}{l|}{\textit{Association Constant} } & $\lambda$                & \multicolumn{2}{c|}{$7\times 10^4$ }    & $M^{-1}s^{-1}$                                    & \multicolumn{1}{c}{}  \\
\hline
\end{tabular}
\caption{Values of physiological parameters for simulations.}
\end{table*}

\subsection{Simulation Results}
In this section, we first present the impulse response simulation and then continue with simulating the effect of mucus flow rate, $v$, ACE2 receptor density, $f(y)$ and mucus thickness, $h(y)$, on the virus-ACE2 binding.
\subsubsection{Impulse Response of Unobstructed Viral Progression}
In Section \ref{impulse_sec}, we find an analytic expression for the impulse response of unobstructed viral progression through the respiratory tract. Here, we confirm our analytic expression with a Monte Carlo simulation in Fig. \ref{fig:impres}.

\begin{figure}
	\centering
	\includegraphics[width=0.5\textwidth]{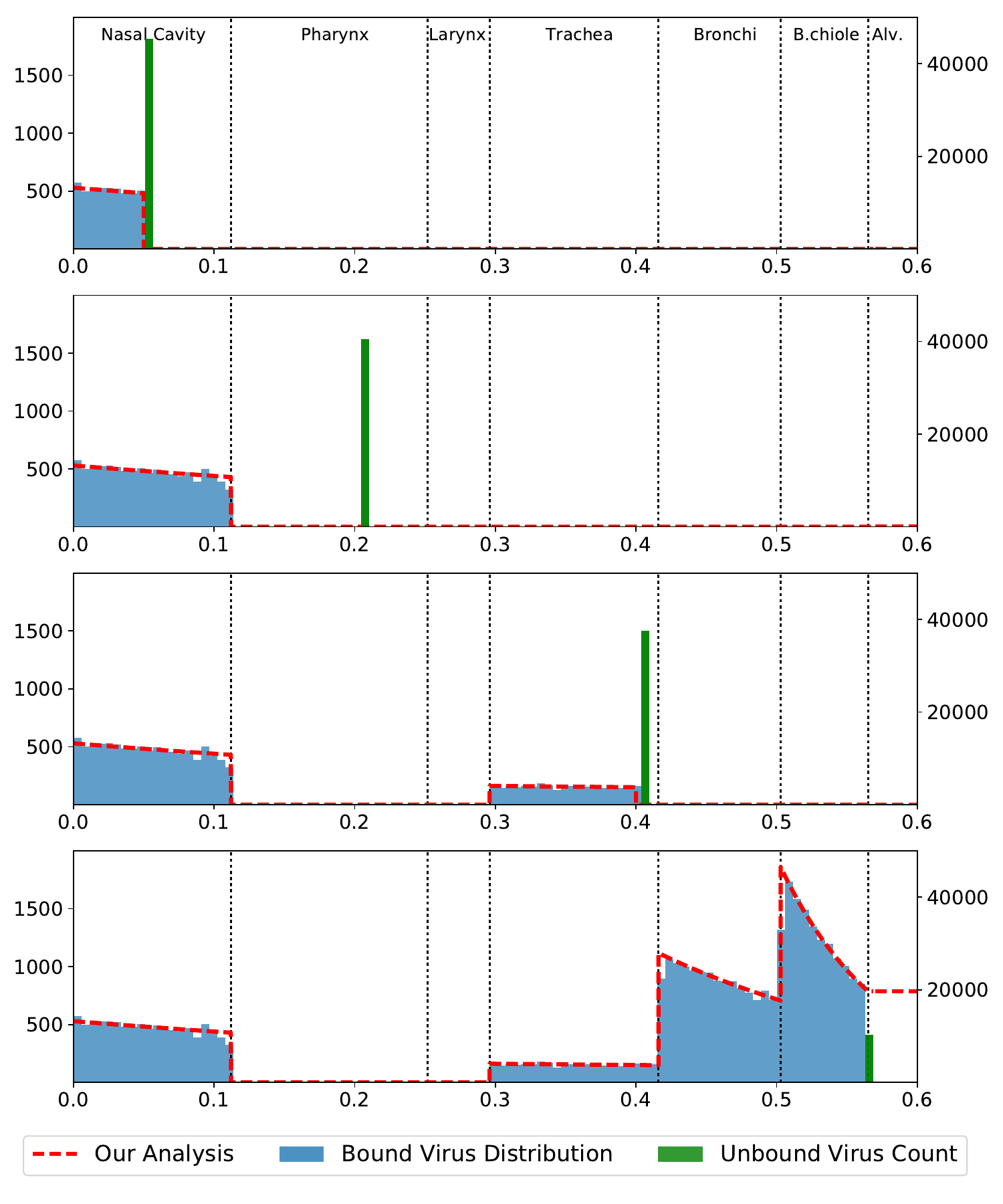}
	\caption{Impulse response of unobstructed viral progression through the respiratory tract for $t=1000s$,$t=4000s$,$t=8000s$ and $t=13000s$ from top to bottom.}
	\label{fig:impres}
\end{figure}

The physiological parameters that we use in the simulations are presented in Table III.
\begin{figure}
	\centering
	\includegraphics[width=0.5\textwidth]{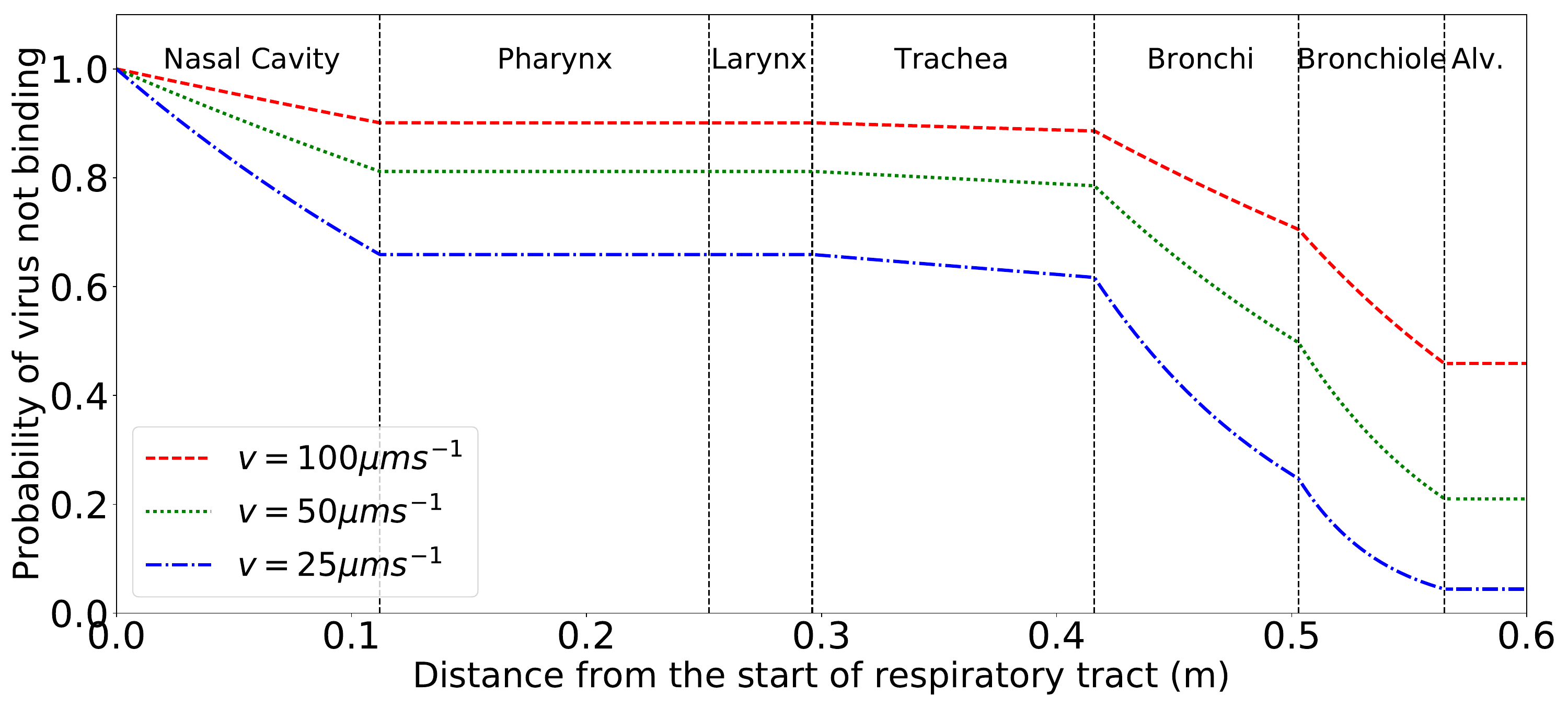}
	\caption{The proportion of the virus population reaching different sections  of the respiratory tract, depending on the mucus flow rate.}
	\label{fig:muc}
\end{figure}

For the Monte Carlo simulation, we divide the respiratory tract to $\Delta{y}=5\mu{m}$ patches. The initial number of viruses are $N_0=50000$. Each virus is independent of each other, i.e., a new number is generated using the Marsenne Twister for each virus in each segment. We see that our analytical solution is in full agreement with the Monte Carlo simulation of the system for large $N_0$.

\subsubsection{Mucus Flow Rate}
As it can be seen in Fig. \ref{fig:muc}, mucus flow rate has a significant impact on the reach of the virus population. If the patient suffers from another condition causing nasal drip or any other faster mucus flow, the virus spends less time in the upper respiratory system. Therefore, ACE2--virus bindings in the upper respiratory tract is limited, causing the bulk of the virus population to migrate to the lower parts of the respiratory tract, especially bronchioles and alveoli. The virus population in the alveoli is 20-folds more if the mucus drop rate is $v=100 \mu{m}s^{-1}$ compared to $v=25 \mu{m}s^{-1}$. This causes the virus to take hold in the alveoli before an immune response can be launched.

\begin{figure}
	\centering
	\includegraphics[width=0.5\textwidth]{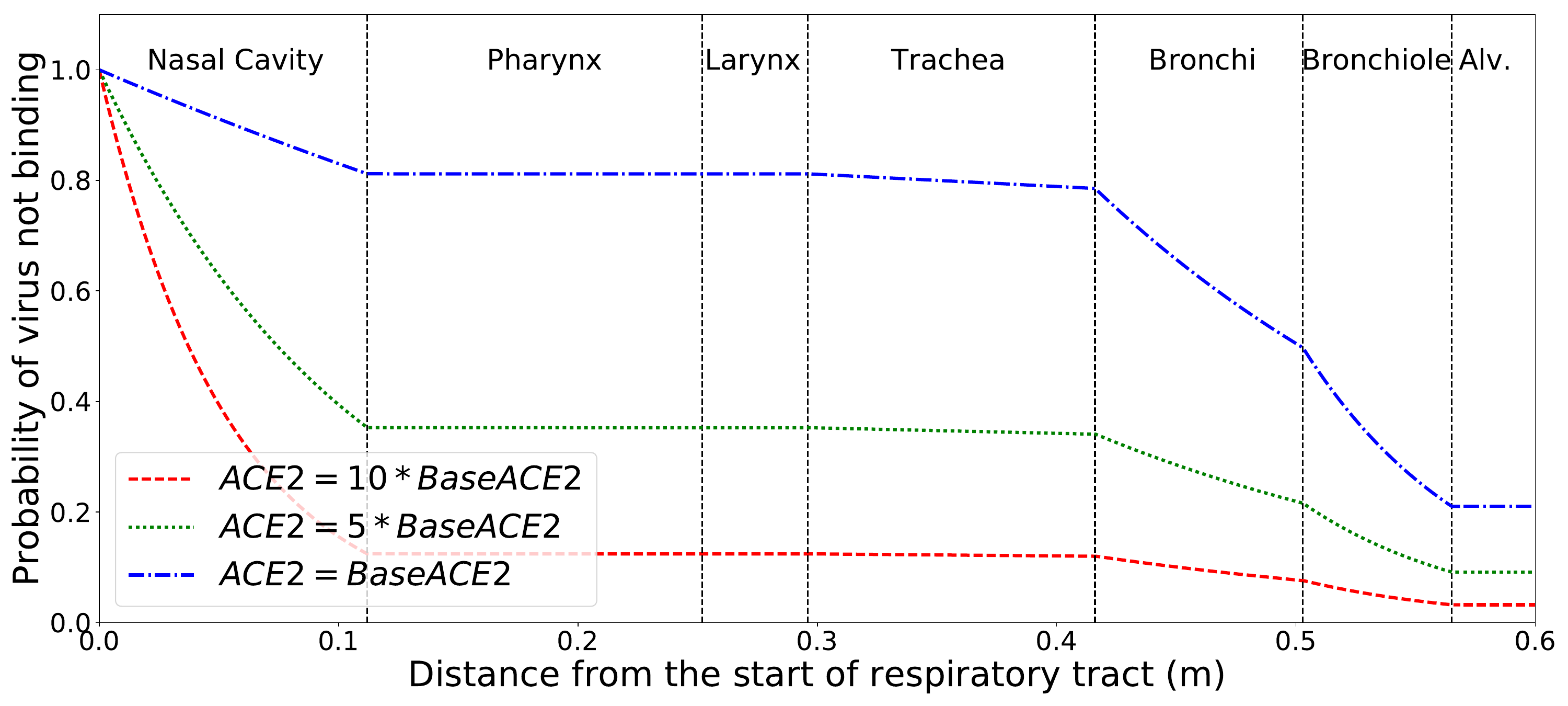}
	\caption{The proportion of the virus population reaching different sections  of the respiratory tract, depending on the ACE2 receptor rate in the nasal cavity.}
	\label{fig:ace}
\end{figure}

\subsubsection{Nasal ACE2 Receptor Density}
Fig. \ref{fig:ace} shows us the impact of the ACE2 receptor concentration in the nasal cavity. Assuming distribution of the ACE2 receptors in the other parts of the respiratory tract is the same for different age groups, the difference in the ACE2 levels in nasal cavity has a significant effect on the virus population reaching the lower respiratory tract. The impact of ten-fold increase in ACE2 receptor concentration is six-fold increase in virus concentration in the lower respiratory system.

\subsubsection{Mucus Thickness}
Our model suggests an impact of the mucus thickness. Since we assume that the virus can move freely in the mucus layer via diffusion, thicker mucus implies that there is less chance for the ACE2--virus binding. Fig. \ref{fig:thick} shows the effect of the mucus thickness. The virus population in the alveoli is 4.45 times more in the four times thicker mucus compared to the base mucus level.

\begin{figure}
	\centering
	\includegraphics[width=0.5\textwidth]{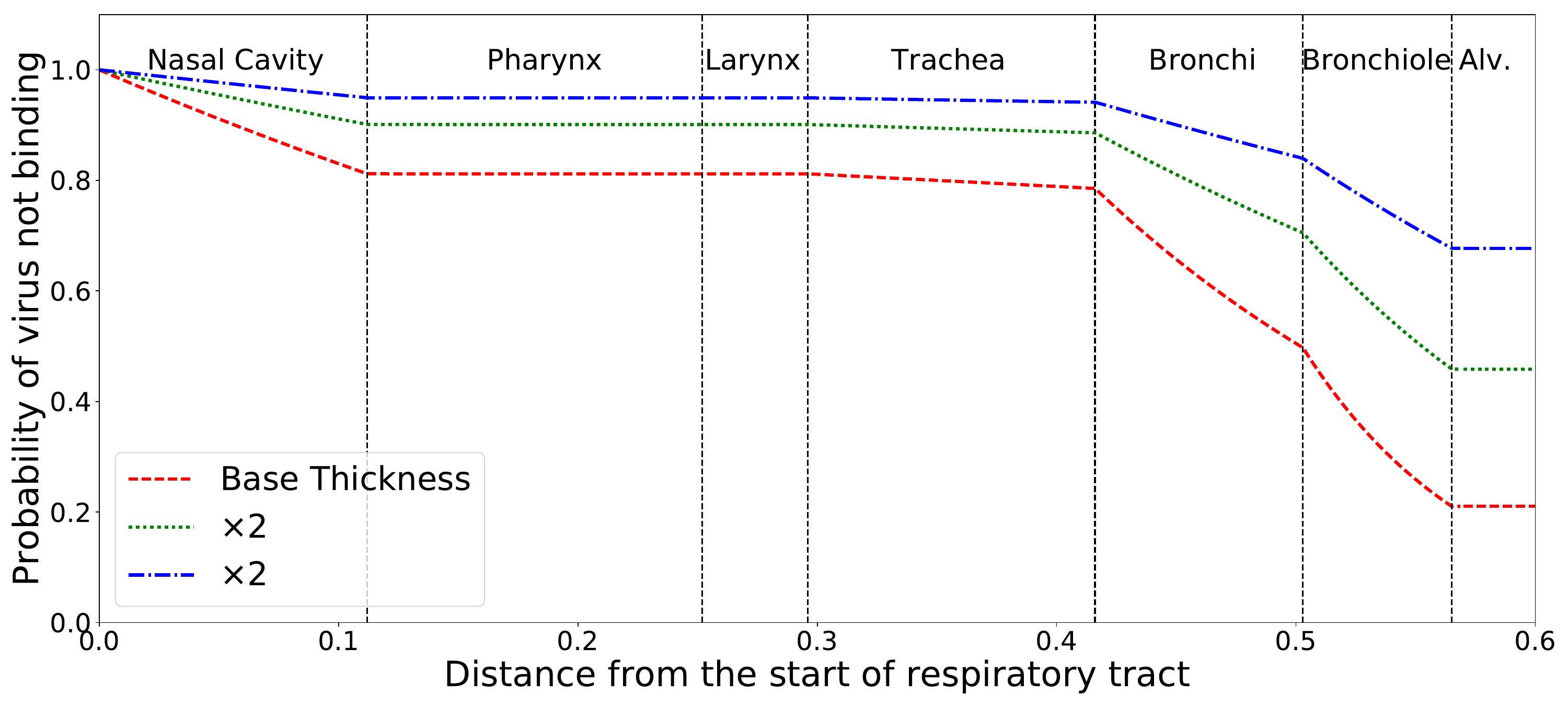}
	\caption{The proportion of the virus population reaching different sections  of the respiratory tract, depending on the mucus thickness in the respiratory tract.}
	\label{fig:thick}
\end{figure}
\section{Conclusion}
\label{concl}
In this work, we use molecular communication to understand the SARS-CoV-2 transmission in human respiratory system. We use a novel respiratory tract model which captures the existing information on ACE2 receptor distribution and mucus flow simultaneously without the computational complexity due to an irregular shape. Our model transforms the respiratory tract into seven consecutive cylinders with different length and radii, each of which has the same ACE2 density and mucus thickness. For the first time in the literature, we calculate the approximate ACE2 receptor densities for all these segments using the existing qualitative and quantitative researches.

Based on the analysis, we reach that higher mucus flow rate results in virus migration to the lower respiratory tract, which is compatible with the experimental results found in the literature. We also analyzed the effects of nasal ACE2 receptor density and mucus thickness. Our findings are aligned with the experimental results in the literature.

Our work will pave the way for more complicated models to analyze the infection mechanisms, especially in terms of viral load necessary for infection, of airborne viruses.

\section{Acknowledgements}
This work was supported  in part by the AXA Research Fund (AXA Chair for Internet of Everything at Koc University), Huawei Graduate Research Scholarship and by Koc University \.{I}\c{s} Bank Research Center for Infectious Diseases (KUISCID).
\bibliographystyle{IEEEtran}
\bibliography{References}

\end{document}